\documentclass[aps,pra,reprint,amsmath,amssymb]{revtex4-1}
\usepackage{graphicx}% Include figure files
\usepackage{dcolumn}% Align table columns on decimal point
\usepackage{bm}% bold math
\begin{document}
%\draft
\title{Metamagnetic jump in %the Magnetization Process \\of  
the spin-$\frac{1}{2}$ 
antiferromagnetic Heisenberg model %\\
on the square-kagome lattice
}
%%%%%%%%%
\author{Yasumasa Hasegawa, Hiroki Nakano, and T\^oru Sakai}
%\address
\affiliation
{Department of Material Science, Graduate School of Material Science, 
University of Hyogo, %\\
3-2-1 Kouto, Kamigori, Hyogo, 678-1297, Japan 
}
%%%%%%
%%%
%\date{Received }
\date{\today}

\begin{abstract}
The magnetization process of the spin-$1/2$ 
antiferromagnetic Heisenberg model %%%_revised_%%%
on two-dimensional square-kagome lattice 
is studied theoretically.
The metamagnetic
jumps exist in the magnetization process
at the higher edge of the $1/3$ and $2/3$ plateaus.  %%%_revised_%%%
% is shown to be caused by the entangled state.
The parameter-dependencies of the critical field and the magnitude of the 
magnetization jump at the higher edge of the $1/3$ plateau  %%%_revised_%%%
are obtained by using the approximated state in the unit cell
and compared with the numerical results of the exact diagonalization
of 42 sites.
\end{abstract}

%\pacs{
%73.22.Pr, %Electronic structure of graphene 
%73.20.-r, %Electron states at surfaces and interfaces
%73.21.-b,
%%Electron states and collective excitations in multilayers, 
%%quantum wells, mesoscopic, and nanoscale systems
%73.40.-c, %Electronic transport in interface structures
%81.05.ue %Graphene
%74.70.Xa, 
%71.10.Fd, 71.27.+a, 74.25.Jb}
\maketitle

\section{Introduction}
\begin{figure}[b]
%
%\flushleft{ (a) \hfill (b) \hfill \ } \vspace{-0.0cm}\\
\begin{center}
%\vspace{1.8cm}
%\vspace*{0.4cm} 
%%%\includegraphics[width=0.33\textwidth]{shurikenlattice2.eps} %\\ \vspace {0.5cm}
\includegraphics[width=0.33\textwidth]{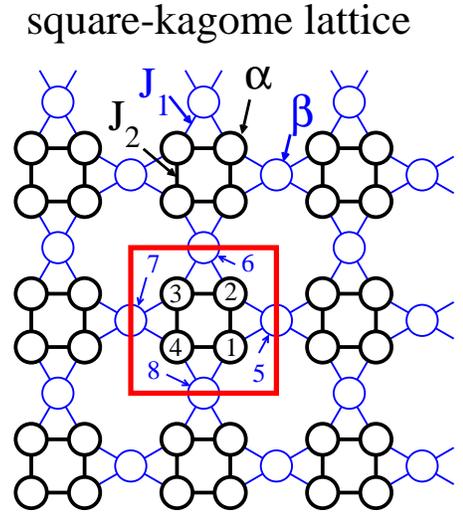} %\\ \vspace {0.5cm}
%\hspace{0.3cm}
%
\end{center}
\caption{(color online).
Square kagome lattice. 
Unit cell is shown by the red square, which consists of four $\alpha$ sites 
($1-4$) and two $\beta$ sites ($5-8$, each spin belongs to two neighboring unit cells), 
forming the \textit{shuriken} structure.}
\label{figshuriken}
\end{figure}

The magnetization process in frustrated Heisenberg spins attracts much interest.
% In the kagome lattice the $7/9$ and $5/9$ plateaus besides $1/3$ plateau are reported\cite{Nishimoto2013}.
%In   the $7/9$ and $5/9$ plateaus, the system is thought to be spontaneously broken the translational symmetry 
%resulting the $\sqrt{3} \times \sqrt{3}$ supercell. 
%Moreover it is shown  that
%the magnetization jump occurs in  the kagome lattice with
 % the $\sqrt{3} \times \sqrt{3}$ modulation\cite{Hida2001}. 
%%%Controversial results on the excitation gaps on the  
%%%antiferromagnetic Heisenberg spins on two dimensional kagome lattice have been reported.
Kagome lattice consists of  triangles and hexagons. 
The triangle structure makes frustration on the system.
Recently, lattice with triangles, squares and octagons, called square kagome lattice
or shuriken lattice (see Fig.~\ref{figshuriken}), has also been 
studied\cite{Siddharthan2001,Rousochatzakis2013,Nakano2013,%
Nakano2014,Nakano2015}. 
It has been reported that besides the magnetic plateaus at $1/3$ and $2/3$ in the magnetization process,
the magnetization jump occurs 
at the high field edge of the $1/3$ plateau\cite{Nakano2013,
Nakano2014,Nakano2015}. 
There exists another magnetization jump 
between $2/3$ 
plateau and the saturation of the magnetization,
which is known to occur in kagome lattice\cite{Schulenburg2002,Zhitomirsky2004}.
Ising spins on the square kagome lattice has also been studied recently.\cite{Pohle2016}
Effective Hamiltonians have been proposed to study the frustrated spin 
systems\cite{Rousochatzakis2013,Bergman2007}.

\begin{figure}[tb]
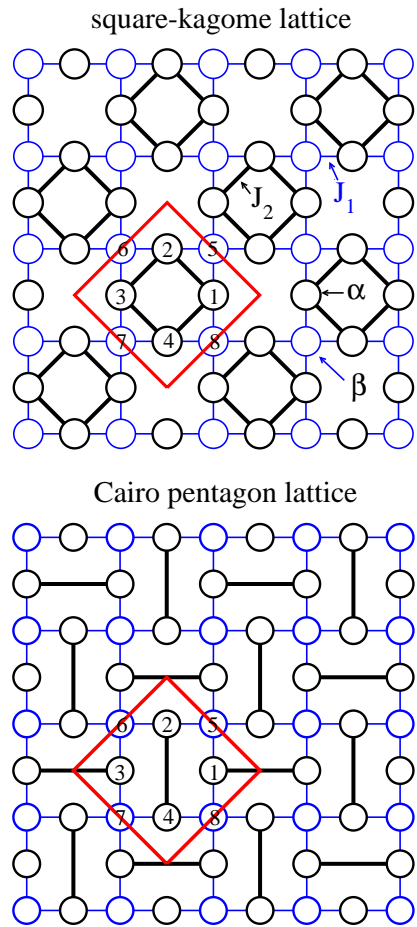

%\flushleft{ (a) \hfill (b) \hfill \ } \vspace{-0.0cm}\\
\begin{center}
%\vspace{1.8cm}
%\vspace*{0.4cm} 
%\includegraphics[width=0.3\textwidth]{shurikenLiebfig.eps} %\\ \vspace {0.5cm}\\
\includegraphics[width=0.3\textwidth]{fig2a.eps} %\\ \vspace {0.5cm}\\
\end{center}
\begin{center}
\includegraphics[width=0.3\textwidth]{fig2b.eps} %\\ \vspace {0.5cm}
%\hspace{0.3cm}
\end{center}
\caption{(color online).
Square kagome lattice, topologically same as Fig.~\ref{figshuriken} and the Cairo pentagon lattice. 
Unit cell is shown by the red square, which consists of four $\alpha$ sites and two $\beta$ sites.}
\label{figshuriken2}
\end{figure}

%The magnetization jumps, or metamagnetic jumps,  have been known 
%as spin flop phenomena in the system in anisotropic spin systems. 
I%n the Heisenberg antiferromagnetic spins on square kagome lattice, on the other hand, the jump occurs
i%n the isotropic spin systems. 
The magnetization jump, or metamagnetic jump in anisotropic spin systems is rather easily understood 
as a spin flop phenomenon, which is a first-order transition between differently ordered states. 
In the Heisenberg antiferromagnetic spins on the square kagome lattice, on the other hand, the jump occurs
in the isotropic spin systems. The magnetization jump  on the square kagome lattice
 is also the first-order transition, but the phases are not so easily imagined as a classical spin picture.
The magnetization jump is also shown to exist
in the square lattice with the next-nearest-neighbor interactions ($J_1-J_2$ model)\cite{Coletta2013},
where the first-order transition between different states occurs.
Recently, another isotropic spin system (Cairo pentagon 
lattice\cite{NakanoIsodaSakai2014,Isoda2014})  has been discovered to have the magnetization jump. 
The Cairo pentagon lattice has no triangle structure but the frustration is caused by the pentagon structure.
The square kagome lattice and the Cairo pentagon lattice can be constructed from the 
Lieb lattice, where frustration does not exist, as shown in Fig.~\ref{figshuriken2}.
It is well known that the Lieb-lattice antiferromagnet 
holds the so-called Marshall-Lieb-Mattis theorem\cite{Marshall,
LiebMattis}. 
This theorem clarifies that this system shows the ferrimagnetic ground state. 
Additional interaction bonds like $J_{2}$ in Fig.~\ref{figshuriken2} 
change the behavior of the system. 
Other types of additional interactions 
were studied\cite{NakanoShimokawaSakai_kgm_dist2011,
NakanoSakai_JJAP2015,Nakano_JJSPM2017}. Among them, 
the kagome-lattice and Lieb-lattice antiferromagnets are connected
by the additional interactions\cite{NakanoShimokawaSakai_kgm_dist2011}. 
There is also another modulation from the kagome-lattice antiferromagnet. 
%Moreover it is shown  that
In the case of the $\sqrt{3} \times \sqrt{3}$ modulation
in the kagome lattice, the magnetization jump  
also occurs\cite{Hida2001, 
NakanoSakai_KgmMag42s2011,Nakano2014}. 
The square kagome lattice  and Cairo pentagon lattice have smaller unit cell
(six spins in the unit cell) than the 
kagome lattice with  the $\sqrt{3} \times \sqrt{3}$ modulation (nine spins in the unit cell), 
so it is more appropriate to study the magnetization jump 
in the frustrated spin systems numerically and analytically. 
    
% 
%%%%%%%%__Fig.1__%%%%%%%%%%%%%%%%%%%%%%%%%%%%%%%%%%%%%%%%%%%%%%%%%%
% 
%%%%%%%%%%%%%%%%%%%%%%%%%%%%%%%%%%%%%%%%%%%%%%%%
    
   Plateau and jump in the magnetization process have also 
been studied in the frustrated Heisenberg 
spin ladder\cite{Honecker2000,Michaud2010}
and in the anisotropic triangular antiferromagnet\cite{Coletta2013}.
It is known that 
the triangular-lattice Heisenberg antiferromagnet shows a plateau 
without jumps at both edges\cite{Bernu_tri1992,Bernu_tri1994,
Lecheminant_tri1995,SakaiNakano_tri_kgm36s2011,Starykh_tri2015}. 
Addition and removal of interactions 
in the triangular-lattice antiferromagnet 
were also studied from the viewpoints 
of the changing plateau behavior\cite{NakanoSakai_tri_nn2017,
NakanoSakai_tri_dice2017,ShimadaNakanoSakaiYoshimura_tri_dice2018}. 
Therefore, it is worth studying how the change of interaction
affects the behavior of various magnetic systems. 

In this paper 
we study the magnetization process,
%%%_added_%%%
especially the $J_2/J_1$ dependencies of the critical magnetic field $h_2$ and
magnitude of the magnetization jump at $h_2$, 
in the square kagome lattice by using the approximated eigenstate
and we give insights for the magnetization jump 
obtained numerically in this system.
%%%_added_%%%
 Rousochatzakis \textit{et al.}\cite{Rousochatzakis2013} have introduced
the effective models  in the square-kagome lattice 
and similar lattices, i.e., the effective interactions between $\beta$ spins around the singlet formed by
four $\alpha$ spins are obtained in the case of $J_1 \ll J_2$. They also
gave the nearest-neighbor valence-bond
description  at $J_1 \approx J_2$, which has been studied 
in the kagome lattice\cite{Singh2007,Poilblanc2010}.
Although they extensively studied the states of $M=0$ and the plateau boundary at $J_2/J_1 \gg 1$,
little attention has been paid to the magnetization jump in the square-kagome lattice at $J_2/J_1 \approx 1$.
%%%_added_%%%  
We show that the magnetization jump 
at the higher edge of the 1/3-plateau 
can be approximated as the uniform phase of the entangled state
%%%_added_%%%
(or the linear combination of the eigenstates)
 %%%_added_%%%
in the unit cell.

\section{square kagome lattice and the exact diagonalization}
    
The square-kagome lattice is shown in Fig.~\ref{figshuriken}. 
There are four $\alpha$ sites and two $\beta$ sites in the unit cell
which is shown by the red square in Fig.~\ref{figshuriken}.  Each $\beta$ site is shared by 
 neighboring unit cells. 
 
% Magnetization plateau in kagome Heisenberg antiferromagnetic 
%is thought as the singlet state of hexagons with $\sqrt{3} \times \sqrt{3}$ period\cite{Capponi2013}

\begin{figure}[bt]
%\flushleft{ (a) \hfill (b) \hfill \ } \vspace{-0.0cm}\\
\begin{center}
%\vspace{1.8cm}
\vspace*{0.4cm} 
\includegraphics[width=0.4\textwidth]{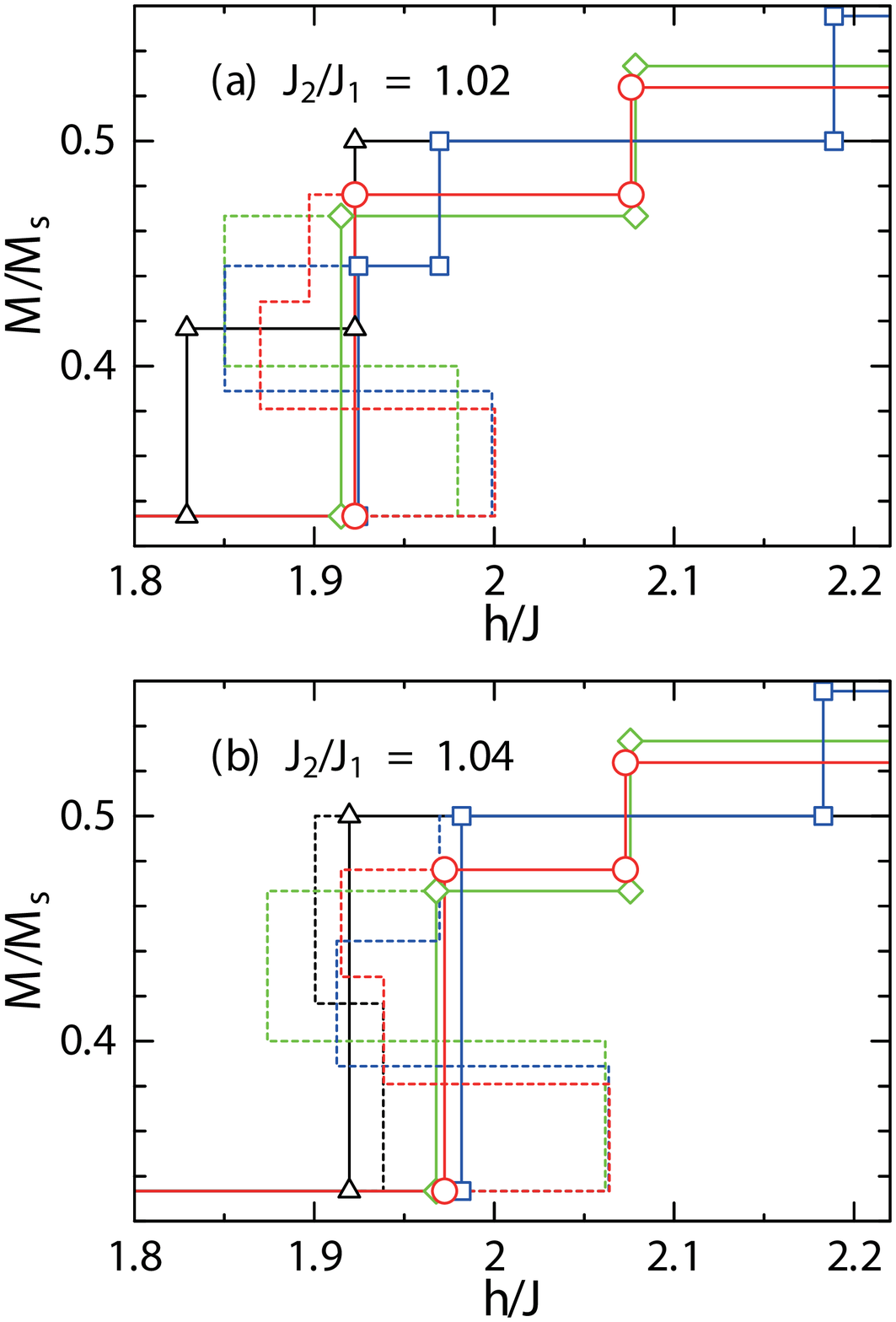} %\\ \vspace {0.5cm}
 %\hspace{0.3cm}
\end{center}
\caption{(color online).
Close-up plot of the  magnetization process near  $1/3$ plateau on square-kagome
 lattice with (a) $J_2/J_1=1.02$ 
and (b) $J_2/J_1=1.04$ obtained by the exact diagonalization.
%%%_added_%%%
The results of $J_2/J_1=1.02$ were already reported  in the previous paper\cite{Nakano2015},
and these of  $J_2/J_1=1.04$ are added 
to study the $J_2/J_1$-dependence on $h_2$.
Black triangles, green diamonds, blue squares, and red circles are obtained in the systems 
of $N_s=24$, $30$, $36$, and $42$, respectively. 
%%%__revised2__%%%%%%%%%%%%%%%%%%%%%
The broken lines represent the results before the Maxwell construction is carried out\cite{Nakano2015}.
%%%__revised2__%%%%%
Magnetization jumps 
are seen in all cases except for the case of $J_2/J_1=1.02$ and  $N_s=24$.
%%%_added_%%%
Note that the critical value $h_2$ depends very little on the size $N_s$ 
at $J_2/J_1=1.02$ and $1.04$ except for $N_s=24$.
}
\label{fignumerical2}
\end{figure}

The Heisenberg model on the square-kagome lattice is given by\cite{Nakano2014}
%%%%%%%%%%%%%%%%%%%%%%
\begin{equation}
\mathcal{H} = \mathcal{H}_1 + \mathcal{H}_2  + \mathcal{H}_{\mathrm{Zeeman}},
 \end{equation}
 where $\mathcal{H}_1$ is the nearest-neighbor interaction between spins on 
 the $\alpha$ and $\beta$ sites,
\begin{equation}
\mathcal{H}_1 = J_1 \sum_{\langle i,j \rangle, i \in \alpha, j \in \beta} \mathbf{S}_i \cdot \mathbf{S}_j , 
\end{equation}
 $\mathcal{H}_2$ is the nearest-neighbor interaction between spins on 
 the $\alpha$ sites,
\begin{equation}
\mathcal{H}_2 = J_2 \sum_{\langle i,j \rangle, i \in \alpha, j \in \alpha} 
\mathbf{S}_i \cdot \mathbf{S}_j ,
\end{equation}
and $\mathcal{H}_{\mathrm{Zeeman}}$ is the Zeeman energy in the magnetic field $h$,
\begin{equation}
\mathcal{H}_{\textrm{Zeeman}}   =-h \sum_{j} S_j^z.
\end{equation}
%%%%%%%%%%%%%%%%
We have reported\cite{Nakano2015} the magnetization process %for the finite systems 
 obtained by exact diagonalization 
in the square kagome lattice
 of $N_s=24$, $30$, $36$, and $42$, where $N_s$ is 
the number of spins and $N_0=N_s/6$ is the number of unit cell.
%
%%%%%%%%__Fig.2__%%%%%%%%%%%%%%%%%%%%%%%%%%%%%%%%%%%%%%%%%%%%%%%%%%
%%%%%%%%%%%%%%%%%%%%%%%%%%%%%%%%%%%%%%%%%%%%%%%%%
%%%%%%%%_fig.3_%%%%%%%%%%%%%%%%%%%%%%%%%%%%%%%%%%%%%%%%%%%%%%%%%%%%%%%%%%%%%%%%
\begin{figure}[tb]
%\flushleft{ (a) \hfill (b) \hfill \ } \vspace{-0.0cm}\\
\begin{center}
%\vspace{1.8cm}
\vspace*{0.4cm} 
\includegraphics[width=0.41\textwidth]{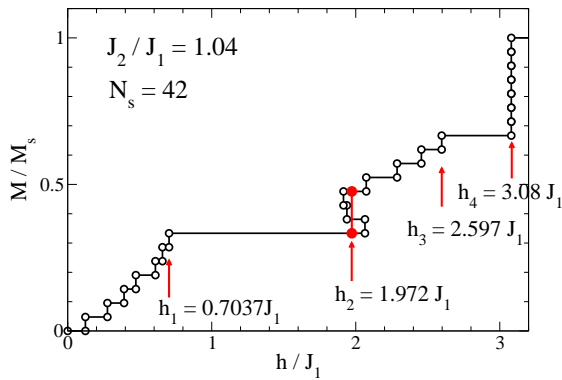}  \vspace {1.0cm} \\
\includegraphics[width=0.4\textwidth]{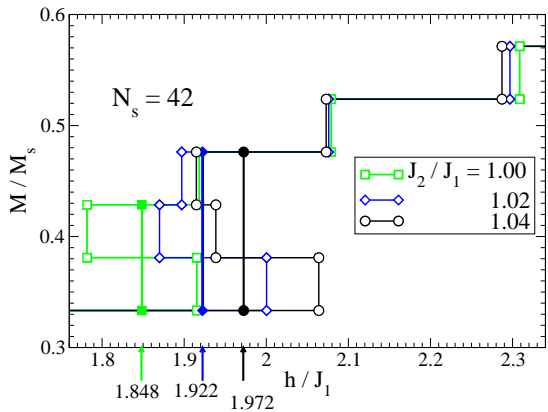}  %\\ \vspace {0.5cm} \\
%\hspace{0.3cm}
%
\end{center}
\caption{(color online).
Magnetization process obtained by the exact diagonalization for the system $J_2/J_1=1.04$ and $N_s=42$
and the close-up plot of the magnetization jump 
for $J_2/J_1=1, 1.02$, and 1.04. The results for $J_2/J_1=1$ and $1.02$ are reported 
in the previous papers\cite{Nakano2013,Nakano2014,Nakano2015},
and the results for $J_2/J_1=1.04$ are the additional data.
The critical fields $h_1$, $h_2$, $h_3$, and $h_4$ are indicated by the red arrows.
The magnetization jumps
are seen at $h_2$ and $h_4$}
\label{figmagpro104}
\end{figure}
%%%%%%%%%%%%%%%%%%%%%%%%%%%%%%%%%%%%%%%%%%%%%%%% 
%
 %%%%%%%%%%%%%%__Fig.4__%%%%%%%%%%%%%%%%%%%%%%%%%%%%%%%%%%%%%%%%%%%%%%%%%%%%%%%%%%%%
\begin{figure}[bt]
%\flushleft{ (a) \hfill (b) \hfill \ } \vspace{-0.0cm}\\
\begin{center}
%\vspace{1.8cm}
\vspace*{0.4cm} 
\includegraphics[width=0.41\textwidth]{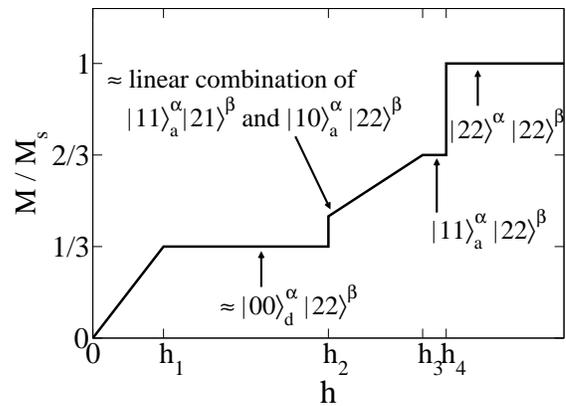} %\\ \vspace {0.5cm}
 %\hspace{0.3cm}
\end{center}
\caption{Schematic figure of the magnetization process of the Heisenberg antiferromagnetic spins 
on square-kagome lattice.
There is $1/3$ plateau at $h_1 \leq h \leq h_2$
and $2/3$ plateau at $h_3 \leq h \leq h_3$.  
The magnetization jumps 
occur at $h=h_2$ and $h=h_4$}
\label{figshurikenmp}
\end{figure}
%%%%%%%%%%%%%%%%%%%%%%%%%%%%%%%%%%
The exact diagonalization is carried out based on the Lanczos algorithm
and the Householder algorithm.
The latter one is used only for the case
when the dimension of the Hilbert space is small.
When the dimension of the Hilbert space becomes extremely large, 
on the other hand,
the Lanczos diagonalization is carried out 
using an MPI-parallelized code, which was originally developed 
in the study of Haldane gaps\cite{NakanoTerai_HaldaneGap2009}. 
The usefulness of our program was confirmed in several large-scale 
parallelized calculations\cite{NakanoSakai_KgmGap2011,
NakanoTodoSakai_s1tri2013,NakanoSakai_KgmMag42s2011,Nakano2015,Nakano2018}. 
The result of the magnetization process obtained by the exact diagonalization
 with the parameters $J_2/J_1 = 1.04$ and $N_s=42$ is shown in Fig.~\ref{figmagpro104}.
 The magnetization process of the Heisenberg antiferromagnetic spins on the square kagome lattice
is shown schematically in Fig.~\ref{figshurikenmp}.
There are plateaus in the magnetization process at $M/M_s=1/3$ and $2/3$ when $h_1 \leq h \leq h_2$ and
 $h_3 \leq h \leq h_4$, respectively. 
 %%%__revised2__%%%%%%%%%%%%%%%%%%%%%%%%%%%%%%%%%%%%%%%%%%%%
 The metamagnetic jump at $h=h_2$  is determined 
by the Maxwell construction\cite{Kohno1997,Nakano2015}
 %%%
 The size dependence of the jump at $h=h_2$ is shown in Fig.~\ref{fignumerical2}. 
The size dependence of $h_2$ is small as obtained from $N_s=30$, 36, and 42.
%
% 
%%%%%%%%%%%%%%%%%%%%%__Fig.5__%%%%%%%%%%%%%%%%%%%%%%%%%%%%%%%%%%%%%%%%%%%
%%%%%%%%%%%%%%%%%%%%%%%%%%%%%%%%%%%%%%%%%%%%%%%%

%%%%%%%%%%%%%%%%%%%%%%%%%%%%%%%%%%%%%%%%%%%%%%%%%%%%%%%%%%%
\section{magnetization process} 
 We define the total spin operators for  $\alpha$ spins (1 - 4 in Fig.~\ref{figshuriken}) 
and  $\beta$ spins (5 - 8 in Fig.~\ref{figshuriken}) as
\begin{equation}
 \mathbf{S}^{\alpha} = \mathbf{S}_1 +  \mathbf{S}_2 + \mathbf{S}_3 + \mathbf{S}_4,
\end{equation} 
and
\begin{equation}
 \mathbf{S}^{\beta} = \mathbf{S}_5 +  \mathbf{S}_6 + \mathbf{S}_7 + \mathbf{S}_8,
\end{equation} 
respectively. 
%%%%__revised2__%%%%
If the system preserves the translational symmetry, $\mathbf{S}_5$ and $\mathbf{S}_6$
 should be the same as $\mathbf{S}_7$ and $\mathbf{S}_8$, respectively. 
Since we are interested in the ground states in the magnetic field 
and the excited states from the plateau states, the translational symmetry may be broken in general.
%
%%%%__revised2__%%%%
Since $\beta$ spins belong to two unit cell simultaneously, the
total spin in the unit cell is given by
\begin{equation}
  \mathbf{S} =  \mathbf{S}^{\alpha}  + \frac{1}{2}  \mathbf{S}^{\beta} 
\end{equation}

When $J_1=0$ we can obtain the 
 eigenstates of the Hamiltonian ($\mathcal{H}_2+\mathcal{H}_{\textrm{Zeeman}}$) as 
shown in Fig.~\ref{figshurikenev0}
(see Appendix~\ref{AppendixA}).
We study the $1/3$ plateau state and the magnetization jump 
 at the higher edge of the magnetic field
($h=h_2$)
by using the approximate states of the entangled state in the unit cell.
We also discuss the jump between $2/3$ plateau and the saturated state at $h=h_4$, 
which is obtained exactly.

%%%%%%%%%%%%%_FIG._6_%%%%%%%%%%%%%%%%%%%%%%%%%%%%%%%%%%%%%%%%%%%%%%%%%%%%%%%%%%%
% 
\begin{figure}[tb]
%
%\flushleft{ (a) \hfill (b) \hfill \ } \vspace{-0.0cm}\\
\begin{center}
%\vspace{1.8cm}
\vspace*{0.4cm} 
\includegraphics[width=0.4\textwidth]{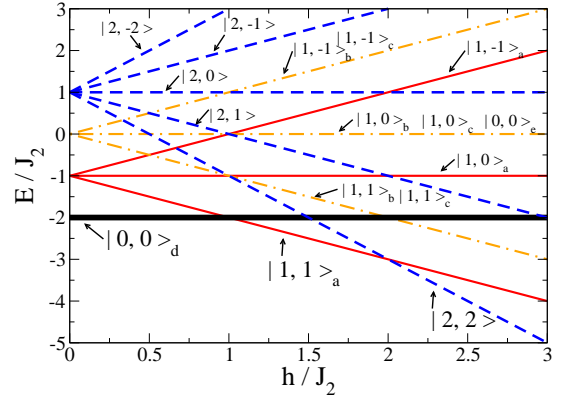} %\\ \vspace {0.5cm}
 %\hspace{0.3cm}
%
\end{center}
\caption{(color online).
Eigenvalues of $\mathcal{H}_2 + \mathcal{H}_{Zeeman}$ of four $\alpha$ 
spins as a function of external magnetic field.}
\label{figshurikenev0}
\end{figure}
%%%%%%%%%%%%%%%%%%%%%%%%%%%%%%%%%%%%%%%%%%%%%%%%
%%%%%%%%%%%

%
 \subsection{$1/3$ plateau state at $h_1 < h < h_2$}
When $J_2 =0$, the square kagome lattice is the same as the Lieb lattice as seen 
in Fig.~\ref{figshuriken2}. 
The Lieb lattice has no frustration.
The ground state of the Lieb lattice with classical spins at $h=0$ is the ferrimagnetic state,  i.e.,
all $\alpha$-spins are up and all $\beta$-spins are down, resulting in the magnetization of $1/3$
of the saturation value. Even if the spins are quantum spins with $S=1/2$, 
the ferrimagnetic state with the $1/3$ magnetization
survives, although the amplitudes of the local spins
are reduced by the quantum effects.

In the other limit  of $J_1=0$ at $h=0$,  
four $\alpha$ spins form 
 the spin singlet state and the $\beta$ spins are arbitrary. 
 %%%%%_added_%%%%
 When $0 < J_1 \ll J_2$,
 the effective interactions between $\beta$ spins have been studied  
by Rousochatzakis \textit{et al.}\cite{Rousochatzakis2013}
 using degenerate perturbation theory.
 They have shown that the ground state at $J_1 \ll J_2$ and $h=0$
can be approximated by the singlet state of four $\alpha$ spins and the crossed-dimer valence bond 
crystal state of $\beta$ spins, resulting in 
 the plateau at $M=0$ due to a finite spin gap. 
 When $J_1 \approx J_2$, the ground states at $h=0$ are different from the ground state
 at $J_1 \ll J_2$ and $h=0$ and 
are not definitely determined\cite{Rousochatzakis2013}.

As we have shown previously\cite{Nakano2015},
the $1/3$ plateau state at $h_1 \leq h \leq h_2$  for $J_2/J_1 \lesssim 0.96$ 
is different from the states for  
$J_2/J_1 \gtrsim 0.96$. 
When $J_2/J_1 \lesssim 0.96 $, the $1/3$ plateau state is the
ferrimagnetic state, similar to that in the Lieb lattice. When $J_2/J_1 \gtrsim 0.96$, the plateau state 
can be approximated by the similar state at  $J_1=0$, i.e.,  the spin singlet state
is formed by four $\alpha$ spins and  all $\beta$ spins
align up. The latter approximation is justified numerically for $J_2/J_1 \gtrsim 1$.
The exact diagonalization studies\cite{Nakano2014, Nakano2015} show
 that in the region of magnetic field $0 < h <h_1$,
 %%%__revised2__%%%%%%%%%%%%%%%%%
$\langle S_i^z \rangle$  ($i \in \beta$) is obtained to be nearly proportional to $h$, 
while $\langle S_i^z \rangle$  ($i \in \alpha$)  is almost zero.
 We approximate   the $1/3$ plateau state as the 
  direct product state of  
$\vert 00 \rangle^{\alpha}_d$ for four $\alpha$ spins and all up states for $\beta$ spins,
i.e.,
 \begin{equation}
 \lvert 0,0 \rangle^{\alpha}_d \vert 2,2 \rangle^{\beta},
 \end{equation}
% where 
%\begin{equation}
 %  \vert 2,2 \rangle^{\beta} = \lvert \uparrow_5 \uparrow_6 \uparrow_7 \uparrow_8 \rangle,
 %\end{equation}
 which can be justified if  $J_2/J_1 \gg 1$.
 Although  the condition $J_2/J_1 \gg 1$
is not fulfilled in the present case, we treat $\mathcal{H}_1$  as perturbation.
%Note that the $\beta$ spins belong to two adjacent
% unit cells simultaneously. Therefore 
The magnetization of this state is 
\begin{equation}
\frac{M}{M_s} %= \frac{0+2}{4+2} 
 =\frac{1}{3}.
\end{equation}
The energy of this state is approximated as
\begin{equation}
 E_{h}^{(h_1 \leq h \leq h_2)} \approx N_0 (-2J_2 -h )
 \label{eqh13}
\end{equation}

\subsection{magnetization jump at $h = h_2$} 
When $h$ is larger than $h_2$, $\alpha$ spins no longer stay a singlet state $\lvert 0,0 \rangle^{\alpha}_d$. 
 In order to increase the magnetization from $1/3$ of the saturation,  four $\alpha$ spins should  become 
 one of the spin-triplet states, which may be
 $\lvert 1, 1\rangle^{\alpha}_a$ state, because this state has the lowest energy when $J_1=0$
 and $J_2<h<2J_2$
(see Fig.~\ref{figshurikenev0}). 
The $z$-component of the $\beta$ spins 
surrounding the $\alpha$ spins may decrease the $z$-component  $S^{\beta}_z$ 
by changing from $\lvert 2,2 \rangle^{\beta}$ to 
$\lvert 2,1\rangle^{\beta}$, as shown in the right figure in Fig.~\ref{figshuriken10a}.
 However, this state is not the eigenstate of $\mathcal{H}_1$ as shown in Appendix~\ref{AppendixD}.
 We approximate the eigenstate just above the magnetic field $h_2$ as a linear combination of
 $\lvert 1 1 \rangle_a^{\alpha} \lvert 21 \rangle^{\beta}$ 
and $\lvert1 0 \rangle_a^{\alpha} \lvert 2 2 \rangle^{\beta}$.
 
 %
%%%%%%%%%%_fig_8-->7%%%%%%%%%%%%%%%%%%%%%%%%%%%%%%%%%%%%%%%%%%%%%%%%%%%%%%%%%%%%%%
\begin{figure}[tb]
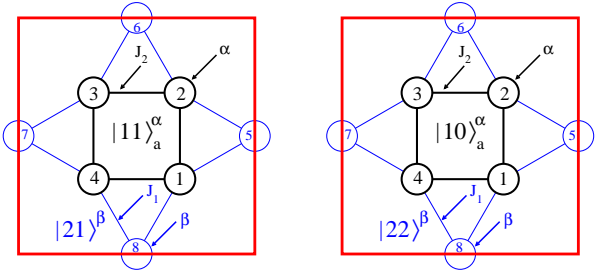

%
%\flushleft{ (a) \hfill (b) \hfill \ } \vspace{-0.0cm}\\
\begin{center}
%\vspace{1.8cm}
%\vspace*{0.4cm} 
%%\includegraphics[width=0.2\textwidth]{shuriken10aa.eps} %\\ \vspace {0.5cm}
\includegraphics[width=0.2\textwidth]{fig8anew.eps} %\\ \vspace {0.5cm} %%%_revised_%%%
  \hspace{5mm}
\includegraphics[width=0.2\textwidth]{fig8bnew.eps} %\\ \vspace {0.5cm} %%%_revised_%%%
 %\hspace{0.3cm}
\end{center}
\caption{Approximate state at $h \gtrsim h_2$ is an entangled state of the two states of
$\lvert11\rangle^{\alpha}_a \lvert21\rangle^{\beta}$ and
$\lvert10\rangle^{\alpha}_a \lvert22\rangle^{\beta}$.}
\label{figshuriken10a}
\end{figure}
%%%%%%%%%%%%%%%%%%%%%%%%%%%%%%%%% 

\begin{figure}[bt]
%\flushleft{ (a) \hfill (b) \hfill \ } \vspace{-0.0cm}\\
\begin{center}
%\vspace{1.8cm}
%\vspace*{0.4cm} 
%%\includegraphics[width=0.43\textwidth]{comp.eps} %\\ \vspace {0.5cm}
\vspace{0.5cm}
\includegraphics[width=0.43\textwidth]{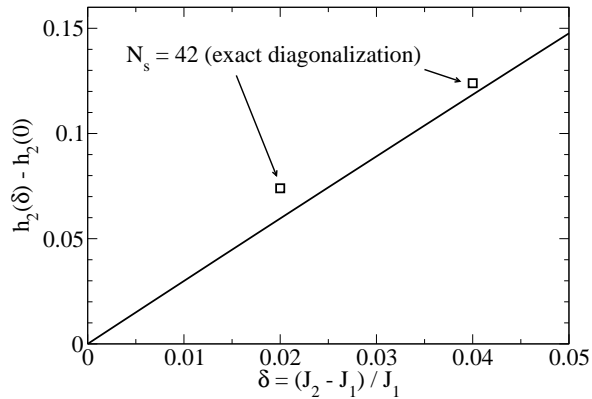} %\\ \vspace {0.5cm}
 %\hspace{0.3cm}
\end{center}
\caption{(color online).
The critical field $h_2$ vs. $(J_2-J_1)/J_1$ given in Eq~\ref{eqh2} (solid line) and the 
numerical results of the exact diagonalization in the system $N_s=42$ (squares).}
\label{figcomp}
\end{figure}

 The state at the  field $h$ just above the higher 
edge of the $1/3$ plateau $h_2$ ($h=h_2+0$) is studied in Appendix~\ref{AppendixD} and  the energy is
 approximately given by
 \begin{equation}
  E_{h}^{(h=h_2+0)} \approx N_0 
  \left( \frac{1}{4} J_1 -J_2 - \frac{5}{4} h - \frac{1}{4} \sqrt{h^2 - 2 J_1 h + 9 J_1^2} \right) .
 \end{equation}
 On the other hand the energy at the $1/3$ plateau is approximated by
Eq.~(\ref{eqh13}).
The upper edge of the 1/3-plateau is obtained by  
\begin{equation} 
 E_{h}^{(h_1 \leq h \leq h_2)} = E_{h}^{(h=h_2+0)}.
 \end{equation}
Then we obtain
\begin{align}
 h_2 &= \frac{(J_1+J_2)(-J_1+2J_2)}{J_2} \nonumber \\
      & = 2 J_1 \frac{(1+\frac{\delta}{2})(1+2\delta)}{1+\delta} 
\label{eqh2} \\
      & \approx 2J_1 \left( 1+\frac{3}{2} \delta \right),
\end{align}
where
\begin{equation}
 \delta = \frac{J_2-J_1}{J_1},
\end{equation}
and we have assumed
\begin{equation}
 0 \leq \delta \ll 1
\end{equation}
Although the absolute value of $h_2$ given in Eq.~\ref{eqh2} 
is a little bit deviated from the value obtained by
the exact diagonalization [$h_2=2$ in Eq.~\ref{eqh2},
 while $h_2=1.848$ is obtained by the exact diagonalization at $J_2=J_1$],
the $(J_2-J_1)/J_1$-dependence of $h_2$ is in good agreement between   Eq.~\ref{eqh2}  and
the exact diagonalization, as shown in Fig.~\ref{figcomp}.
We will discuss the interaction between the excitations in the next section.

At the magnetic field just above $h_2$,
the eigenstate is approximated as
\begin{equation}
 \lvert \Psi \rangle \approx \frac{-\sqrt{2}J_2}{\sqrt{J_1^2+2 J_2^2}} 
\lvert 11\rangle_a^{\alpha} \lvert 21 \rangle^{\beta} 
     + \frac{J_1}{\sqrt{J_1^2+2 J_2^2}} \lvert10 \rangle_a^{\alpha} \lvert 22 \rangle^{\beta}.
\end{equation}
 The magnetization at $h =h_2+0$ is
 \begin{align}
  \langle m \rangle &= \frac{1}{J_1^2+2 J_2^2} \left( \frac{1}{3} J_1^2 + J_2^2\right) \nonumber \\
   &= \frac{4\left( 1+\frac{3}{2}\delta +\frac{3}{4}\delta^2 \right)}%
{9\left(1+\frac{4}{3}\delta + \frac{2}{3} \delta^2\right)} \nonumber \\
   & \approx \frac{4}{9} \left( 1+\frac{1}{6} \delta \right).
   \label{eqmagh2p}
 \end{align}
%If the excitations from the state at the 1/3-plateau attract each other,
%the magnetization jumps 
%from %1/2 
%$1/3$
%to the value given in Eq.~(\ref{eqmagh2p}).
%
%
%
%

%%%%__revised2__%%%%%%%%%%%%%%%%%%%%%%%%%%%%%%
If the interaction between the excitations from the state at the $1/3$ state were repulsive,  
the Bose-Einstein condensation of magnons would
happen, which has been shown to be realized in 
several materials.\cite{Nikuni2000,Rice2002,Jaime2004,Samulon2009, Zapf2014} 
In that case
the magnetization would increase continuously when magnetic field is increased from $h_2$.
However, as we will show numerically in the next section, the interaction between the 
excitations from the state at the $1/3$ state is attractive. Then the excitations occur on every unit cell 
in the ground state at $h=h_2+0$. 
In this case the magnetization jumps from $1/3$ to the value given in Eq.~(\ref{eqmagh2p}).
%%%%__revised2__%%%
%

 %
\subsection{$2/3$ plateau at $h_3 < h < h_4$ and the jump at $h=h_4$} 
We study the $2/3$ magnetization plateau and  jump at $h=h_4$ in this subsection
in order to make clear the mechanism of the jump in this system.
All spins align to the $z$ direction at $h>h_4$. 
This state is written as the direct product of the $S^{\alpha}=2$, $S_z^{\alpha}=2$ 
state of four $\alpha$ spins
($\lvert 2,2 \rangle^{\alpha}$) and the $S^{\beta}=2$, $S_z^{\beta}=2$ state 
of four $\beta$ spins ($\lvert 2,2 \rangle^{\beta}$).
We write the state at $h > h_4$ as
\begin{equation}
 \lvert 2,2 \rangle^{\alpha} \lvert 2,2 \rangle^{\beta}
\end{equation}
%Note that each $\beta$ spin belongs to two unit cells.  
The magnetization per unit cell is $M_s/N_0=3$ ($M/M^s$=1) 
and  the energy per unit cell is obtained as
\begin{equation}
 \frac{E_{h}^{(h \geq h_4)}}{N_0} = ( J_2 + 2 J_1 - 3h ) .% \ \ (h \geq h_4).
 \label{eqeneh4}
\end{equation}

%%%%%%%%%%%%%%%%%%%%%__Fig._9-->8%%%%%%%%%%%%%%%%%%%%%%%%%%%%%%%%%%%%%%%%%%%
% 
\begin{figure}[tb]
%
%\flushleft{ (a) \hfill (b) \hfill \ } \vspace{-0.0cm}\\
\begin{center}
%\vspace{1.8cm}
\vspace*{0.4cm} 
\includegraphics[width=0.33\textwidth]{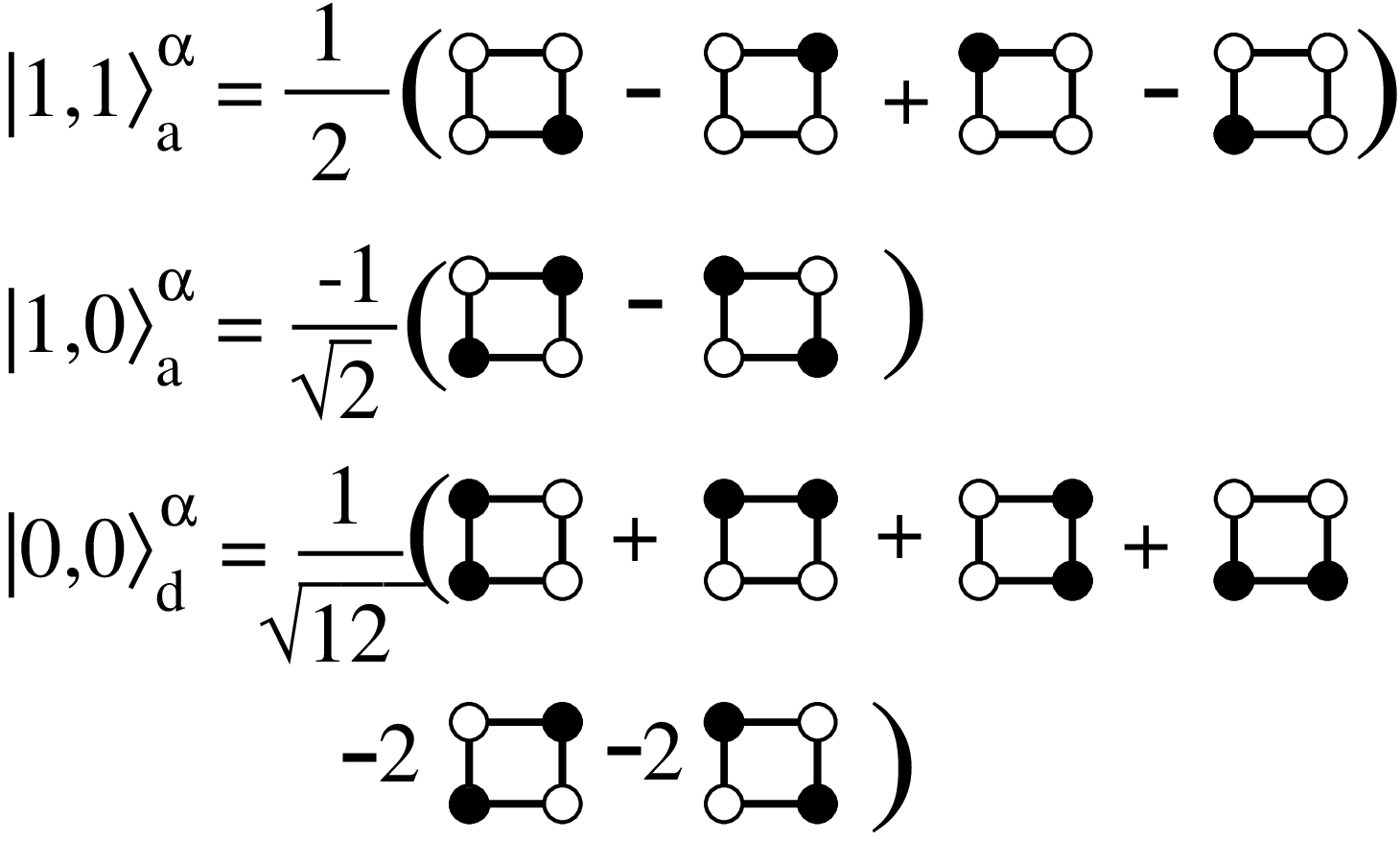} %\\ \vspace {0.5cm}
 %\hspace{0.3cm}
%
\end{center}
\caption{(color online).
Some eigenstates of $\mathcal{H}_2 + \mathcal{H}_{\textrm{Zeeman}}$. 
Open circles are up spins and filled circles are down spins at cites $1, 2, 3$, and $4$.%
}
\label{figqstates}
\end{figure}
%%%%%%%%%%%%%%%%%%%%%%%%%%%%%%%%%%%%%%%%%%%%%%%%
%
%%%%%%%%__Fig.7_-->9%%%%%%%%%%%%%%%%%%%%%%%%%%%%%%%%%%%%%%%%%%%%%%%%%%
%%%%%%%%%%%%%%%%%%%%%%%%%%%%%%%%%%%%%%%%%%%%%%%%
When we decrease the magnetic field below $h_4$, the magnetization jumps  
from the fully polarized state ($M/M_s=1$)
to the $2/3$ plateau. This jump can be understood as follows. 
In this $2/3$
plateau the spins at the $\beta$ sites are aligned to the $z$ direction, 
while the four $\alpha$ spins form the 
spin triplet $\lvert 1,1\rangle_a^{\alpha}$, 
since $\lvert 1,1\rangle_a^{\alpha}$ is the lowest state within 
$S=1, S_z=1$ states for four $\alpha$ spins (See Fig.~\ref{figshurikenev0}). 
 Note that both  $\lvert 2,2\rangle^{\alpha} \lvert 2,2\rangle^{\beta}$
and $\lvert 1,1\rangle_a^{\alpha} \lvert 2,2\rangle^{\beta}$ 
are eigenstates of the Hamiltonian with the energy given as Eq.~(\ref{eqeneh4})
and
\begin{equation}
 \frac{E_{h}^{(h_3 \leq h \leq h_4)}}{N_0} = ( - J_2 +  J_1 - 2h ) ,
\end{equation}  
respectively.
In both states the shared $\beta$ spins are all up's. 
Therefore,  any spatially mixed states of 
 $\lvert 2,2\rangle^{\alpha} \lvert 2,2\rangle^{\beta}$
and $\lvert 1,1\rangle_a^{\alpha} \lvert 2,2\rangle^{\beta}$ are
also the eigenstates. 
If the fraction of $p$ ($0 \leq p \leq 1$) of the
unit cells is the state
$\lvert 2,2\rangle^{\alpha} \lvert 2,2\rangle^{\beta}$ and
$(1-p)$ of the unit cells is the state
$\lvert 1,1\rangle_a^{\alpha} \lvert 2,2\rangle^{\beta}$,
the energy is 
\begin{equation}
 \frac{E}{N_0} = p  \frac{E_{h}^{(h \geq h_4)}}{N_0}   +(1-p) \frac{E_{h}^{(h_3 \leq h \leq h_4)}}{N_0} .
\end{equation}
The lowest energy is obtained by $p=0$ for $h<h_4$ and by $p=1$ for $h>h_4$, where
the critical value, $h_4$, is obtained by the equation
\begin{equation}
 E_{h_4}^{(h \geq h_4)}= E_{h_4}^{(h_3 \leq h \leq h_4)}.
 \end{equation}
 We obtain 
 \begin{equation}
  h_4 = J_1+2 J_2.
 \end{equation}
In this subsection no approximation is used.
A similar situation has been studied for the magnetization jump 
to the saturated magnetization in kagome lattice\cite{Schulenburg2002,Zhitomirsky2004}

 \section{Interaction between excitations}
 %%%%%%%%%%%_fig_10_%%%%%%%%%%%%%%%%%%%%%%%%%%%%%%%%%%%%%%%%%%%%%%%%%%%%%%%%%%%%%
% 
\begin{figure}[tb]
%
%\flushleft{ (a) \hfill (b) \hfill \ } \vspace{-0.0cm}\\
\begin{center}
%\vspace{1.8cm}
\vspace*{0.4cm} 
\includegraphics[width=0.4\textwidth]{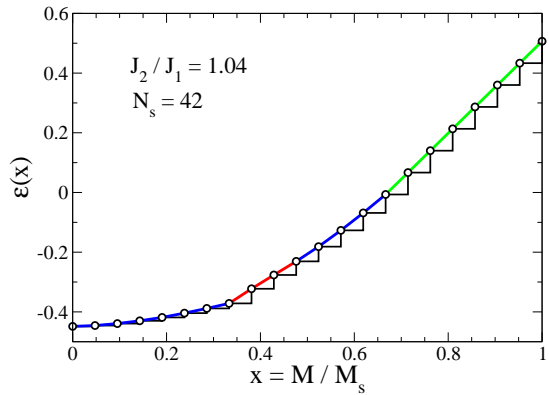} \\ %\\ \vspace {0.5cm} \\
\vspace*{0.9cm} 
\includegraphics[width=0.4\textwidth]{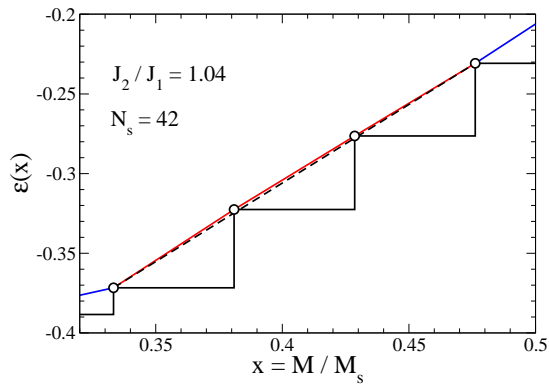} %\\ \vspace {0.5cm}
 %\hspace{0.3cm}
%
\end{center}
\caption{(color online).
Energy per site [$\epsilon(x)$] as a function of $x = M/M_s=n/21$ obtained by the exact diagonalization
($N_s=42$).
At $0 < x \leq 1/3$, $\epsilon(x)$ is downward convex as shown by the blue line. 
A kink is seen at $x=1/3$. At $1/3 \leq x \leq 10/21$ the curvature is upward convex (the red line) and 
a little higher than the black broken line connecting the third-nearest circles at 
 $x=1/3=7/21$ and $x=10/21$. At $10/21 \leq x \leq 2/3$, $\epsilon(x)$ is again downward convex
(the blue line). At $x=2/3$ there is a kink again, and
 all circles at $2/3 \leq x \leq 1$ are on the straight green line.
 }
\label{figenevsx104}
\end{figure}
%%%%%%%%%%%%%%%%%%%%%%%%%%%%%%%%%%%%%%%%%%%%%%%%
%
In this section we consider the interaction between excitations.
 We take $x=M/M_s$, where $M$ is the magnetization and $M_s$ is 
the saturation value of the magnetization, $M_s=N_s/2$ ($N_s$ is the number of sites).
We define the energy $E(x)$ as the lowest energy at $h=0$ (the eigenvalue of
$\mathcal{H}_1+\mathcal{H}_2$) among the eigenstates 
having the same magnetization $x$.
  In the finite system, $x$ can have the discrete values
\begin{equation}
 x =  \frac{2 n}{N_s},
\end{equation}
where 
\begin{equation}
 n = 0, \pm 1, \pm2, \cdots, \pm \frac{N_s}{2}.
\end{equation}
We have assumed that $N_s$ is an even number.
We define the lowest energy per site at $h=0$ among the states with the magnetization $x$, 
\begin{equation}
\epsilon(x) = \frac{E(x)}{N_s}.
\end{equation}
In Fig.~\ref{figenevsx104} we plot  $\epsilon(x) $ as a function of $x$ 
obtained by the exact diagonalization with $J_2/J_1=1.04$ and $N_s=36$.
The magnetization process is calculated as
\begin{align}
 h(x) &= E(x)-E\left(x-\frac{2}{N_s}\right) \nonumber \\
       &=N_s \left[ \epsilon(x) - \epsilon\left(x-\frac{2}{N_s}\right) \right].
\end{align}
In the limit of $N_s \to \infty$, it becomes
\begin{equation}
 h(x) = 2 \frac{d \epsilon(x)}{ d x}.
\end{equation}
In the magnetization process, we plot the 
 magnetization $x$ as a function of $h$, as shown in  Fig.~\ref{figmagpro104}.
 
%ToBeContinued
In the region $0<x \leq 1/3$ ($0<h \leq h_1$), the graph of $\epsilon(x)$ is downward convex
(blue lines).
This downward convex curvature means the repulsive interaction between the 
magnon-like excitations from the totally singlet state at $h=0$.
The plateau at $x=1/3$ corresponds to the kink of $\epsilon(x)$ at $x=1/3$.
Above $x=1/3$  the graph of $\epsilon(x)$ is upward convex (red lines) as shown in Fig.~\ref{figenevsx104}.
%%%_added
Although the difference between the red lines connecting the nearest circles and the black broken line
connecting the third-nearest circles is very small, it is much larger than the  
numerical errors of the exact diagonalization (relative errors should be less than $10^{-10}$ for example).
The downward convex curvature means that the attractive interaction works between the excitation
from the plateau state. Thus the entangled states studied approximately in the previous section 
will be created in all unit cells, resulting in the finite jump in the magnetization.
The straight line of $\epsilon(x)$ at $2/3 \leq x \leq 1$ is consistent with no-interaction 
between the excitation from the $2/3$ state or from the fully saturated state.

Finally, we would like to comment on the experimental situation. 
Although a good candidate material for the present system 
depicted in Fig.~\ref{figshuriken}, 
%a similar material 
was reported\cite{Fujihala_JPS_Iwate2017} and the numerical study was done\cite{Morita2018}, 
%in which 
the material has a further additional distortion.% $J_{3}$ is included.
Owing to this addition, the behavior of this material is different 
from the present result\cite{Morita2018}. 
Even though there is such a difference, 
good candidate materials will be found in the near future. 

 %%%%%%%%%%%%%%%%%%%%%%%%%%%%%%%%%%%%%%%
 \section{conclusion}
 In this paper we study the magnetization process of the Heisenberg
 anti-ferromagnet on the square kagome lattice
 by using the approximated wave function. 
We take the approximation that the ground state just above the higher edge of the $1/3$ plateau
is the entangled state of the $S=1$
 triplet states of the $\alpha$ spins on the square and  the $S=2$ quintet state of the
$\beta$ spins, $\lvert10\rangle^{\alpha}_a \lvert22\rangle^{\beta}$ and
$\lvert11\rangle^{\alpha}_a \lvert21\rangle^{\beta}$.
 Since the $\beta$ spins are shared by neighboring 
unit cells, the magnetization of the entangled states depend on the coefficient of two states.
 In spite of the crude approximation taken in this paper, 
 it gives the reasonable $J_2/J_1$ dependence of the value of the critical field $h_2$ and the
magnitude of the magnetization jump, 
which are obtained by the exact diagonalization study.
%%%%%_added
The approximation is justified when $J_2/J_1 \gg 1$. The reason it 
seems to work well 
even when $J_2/J_1 \gtrsim 1$ would be the frustration, which reduces the
effective coupling ($J_1$) between the spins forming triangles with respect to the coupling
($J_2$) in the spins forming squares without frustrations. 
%%%%
\begin{acknowledgments}
This work was partly supported by JSPS KAKENHI Grant No. 
16K05418, No. 16K05419, and No. 16H01080(JPhysics). 
Nonhybrid thread-parallel calculations in numerical diagonalizations 
were based on TITPACK version 2 coded by H. Nishimori. 
In this research, we used the computational resources 
of the K computer provided by the RIKEN Advanced 
Institute for Computational Science through the HPCI 
System Research projects (Project ID: hp170018, 
hp170028, and hp170070). We used the computational 
resources of Oakforest-PACS provided by Joint Center 
for Advanced High Performance Computing 
through the HPCI System Research project (Project ID hp170207).
%%
%%This research used computational resources of the K computer 
%%provided by the RIKEN Advanced Institute for Computational Science 
%%through the HPCI System Research projects 
%%(Project ID: hp170017, hp170028, hp170070, and hp170207). 
%%
Some of the computations were performed using facilities 
of the Department of Simulation Science, 
National Institute for Fusion Science; Institute for Solid State Physics, 
The University of Tokyo; and Supercomputing Division, 
Information Technology Center, The University of Tokyo. 
This work was partly supported by the Strategic Programs 
for Innovative Research; the Ministry of Education, Culture, 
Sports, Science and Technology of Japan; 
and the Computational Materials Science Initiative, Japan.
%In this research, we used the
%computational resources of the K computer provided by the RIKEN Advanced
%Institute for Computational Science through the HPCI System Research projects
%(Project ID: hp170028).
 \end{acknowledgments}

% \clearpage
 \appendix
 \section{eigenstates of $\mathcal{H}_2 + \mathcal{H}_{\mathrm{Zeeman}}$}
 \label{AppendixA}
 The eigenstates of $\mathcal{H}_2$ for four $\alpha$ spins on the corner of the square 
is written as the linear combination of
 $ \lvert \sigma_1, \sigma_2, \sigma_3, \sigma_4 \rangle$, where
 $\sigma_j = \uparrow$ or $\downarrow$ (see Fig.~\ref{figqstates}). 
The eigenstates are also written as $ \lvert S, S_z \rangle^{\alpha}$,
 where $S$ is the total spin for four spins on $\alpha$ sites and $S_z$ is the $z$ component of total spin.
%\begin{equation}
% \mathbf{S}^{\alpha} = \mathbf{S}_1 + \mathbf{S}_2 + \mathbf{S}_3 + \mathbf{S}_4.
%\end{equation} 
The same notations are used for the four $\beta$ spins ($5, 6, 7$, and $8$). 

 There are one $S=2$ quintet, three $S=1$ triplets and two $S=0$ singlets.
%We take the simultaneous eigenstates for $\mathbf{S}$, $S_z$ and $\mathcal{H}_2$.
 The $S=2$ states are given as
 \begin{align}
   | 2,2 \rangle^{\alpha} &=                         \lvert \uparrow   \uparrow \uparrow \uparrow \rangle \\
   | 2,1 \rangle^{\alpha} &= \frac{1}{2} (       \lvert \downarrow \uparrow \uparrow \uparrow \rangle 
                                     +            \lvert \uparrow    \downarrow \uparrow \uparrow \rangle \nonumber \\
                     &  \hspace{5mm} +  \lvert \uparrow    \uparrow \downarrow \uparrow \rangle 
                                     +       \lvert \uparrow   \uparrow \uparrow \downarrow \rangle         )\\
   | 2,0 \rangle^{\alpha} &=\frac{1}{\sqrt{6}} (  \lvert \uparrow   \uparrow \downarrow \downarrow \rangle 
                                      +           \lvert \uparrow  \downarrow \downarrow \uparrow \rangle  \nonumber \\
                    &\hspace{5mm}   +    \lvert \downarrow \downarrow \uparrow \uparrow \rangle 
                    +                             \lvert \downarrow \uparrow \uparrow \downarrow \rangle  \nonumber \\
                &\hspace{5mm}        +   \lvert \uparrow    \downarrow \uparrow \downarrow \rangle 
                                  +                \lvert \downarrow \uparrow \downarrow \uparrow \rangle )\\
   | 2,-1 \rangle^{\alpha} &= \frac{1}{2} (           \lvert \uparrow \downarrow \downarrow \downarrow \rangle
                                    +              \lvert \downarrow \uparrow \downarrow \downarrow \rangle  \nonumber \\
                    &  \hspace{5mm} +     \lvert \downarrow \downarrow \uparrow \downarrow \rangle
                                    +      \lvert \downarrow \downarrow \downarrow \uparrow \rangle 
       ) \\
  | 2,-2 \rangle^{\alpha} &=         \lvert \downarrow \downarrow  \downarrow \downarrow \rangle                  
\end{align}
%%%%

We write the three $S=1$ triplets 
as $\lvert 1, S_z \rangle_a $,  $\lvert 1, S_z \rangle_b$, and $\lvert 1, S_z \rangle_c $,
which are given by
\begin{align}
 \lvert 1,1 \rangle_a^{\alpha} &= \frac{1}{2} ( \lvert \downarrow \uparrow \uparrow \uparrow \rangle
                                               -  \lvert \uparrow \downarrow \uparrow \uparrow \rangle \nonumber \\
                       & \hspace{5mm} +\lvert \uparrow \uparrow \downarrow \uparrow \rangle
                                               -  \lvert \uparrow \uparrow \uparrow \downarrow \rangle   )  \\
 \lvert 1,0 \rangle_a^{\alpha} &= \frac{1}{\sqrt{2}} ( - \lvert \uparrow \downarrow \uparrow \downarrow \rangle
                                               +  \lvert \downarrow \uparrow \downarrow \uparrow \rangle ) \\
 \lvert 1,-1 \rangle_a^{\alpha} &= \frac{1}{2} ( -\lvert \uparrow \downarrow \downarrow \downarrow \rangle
                                               +  \lvert \downarrow \uparrow \downarrow \downarrow \rangle \nonumber \\
                       & \hspace{5mm} -\lvert \downarrow \downarrow \uparrow \downarrow \rangle
                                               +  \lvert \downarrow \downarrow \downarrow \uparrow \rangle   )         
 \end{align}
 \begin{align}
\lvert 1,1 \rangle_b^{\alpha} &=\frac{1}{\sqrt{2}} (\lvert \downarrow \uparrow \uparrow \uparrow \rangle
                                               -  \lvert \uparrow \uparrow \downarrow \uparrow \rangle ) \\                                    
\lvert 1,0 \rangle_b^{\alpha} &= \frac{1}{2} (- \lvert \uparrow \uparrow \downarrow \downarrow \rangle
                                               -  \lvert \uparrow \downarrow \downarrow \uparrow \rangle  \nonumber \\
                     & \hspace{5mm} + \lvert \downarrow \downarrow \uparrow \uparrow \rangle
                                               +  \lvert \downarrow \uparrow \uparrow \downarrow \rangle  )  \\
\lvert 1,-1 \rangle_b^{\alpha} &=\frac{1}{\sqrt{2}}(-\lvert \uparrow \downarrow \downarrow \downarrow \rangle
                                               +  \lvert \downarrow \downarrow \uparrow \downarrow \rangle   ) 
 \end{align}
 \begin{align} 
\lvert 1,1 \rangle_c^{\alpha} &=\frac{1}{\sqrt{2}} (\lvert \uparrow \downarrow \uparrow \uparrow \rangle
                                               -  \lvert \uparrow \uparrow \uparrow \downarrow \rangle ) \\          
\lvert 1,0 \rangle_c^{\alpha} &= \frac{1}{2} ( -\lvert \uparrow \uparrow \downarrow \downarrow \rangle
                                               + \lvert \uparrow \downarrow \downarrow \uparrow \rangle  \nonumber \\
                     & \hspace{5mm} +  \lvert \downarrow \downarrow \uparrow \uparrow \rangle
                                               -  \lvert \downarrow \uparrow \uparrow \downarrow \rangle  )  \\       
\lvert 1,-1 \rangle_c^{\alpha} &=\frac{1}{\sqrt{2}}(- \lvert \downarrow \uparrow \downarrow \downarrow \rangle
                                               +  \lvert \downarrow \downarrow \downarrow \uparrow \rangle )      
\end{align}

We write  two $S=0$ singlets  as $\lvert 0,0 \rangle_d$ and $\lvert 0,0 \rangle_e$, which are given by
\begin{align}
 \lvert 0,0 \rangle_d^{\alpha}  &= \frac{1}{\sqrt{12}} (
                              \lvert \uparrow \uparrow \downarrow \downarrow \rangle 
                           +  \lvert \uparrow \downarrow \downarrow \uparrow \rangle \nonumber \\
   & \hspace{5mm} +  \lvert \downarrow \downarrow \uparrow \uparrow \rangle 
                            + \lvert \downarrow \uparrow \uparrow \downarrow \rangle  \nonumber \\
     & \hspace{5mm} -2  \lvert \uparrow \downarrow \uparrow \downarrow \rangle
                             -2  \lvert \downarrow \uparrow \downarrow \uparrow \rangle       ) \\
  \lvert 0,0 \rangle_e^{\alpha}  &= \frac{1}{2} (    
                             \lvert \uparrow \uparrow \downarrow \downarrow \rangle 
                           - \lvert \uparrow \downarrow \downarrow \uparrow \rangle \nonumber \\
   & \hspace{5mm} +  \lvert \downarrow \downarrow \uparrow \uparrow \rangle 
                            - \lvert \downarrow \uparrow \uparrow \downarrow \rangle  )              
\end{align} 

The $S=2$ states are
the  eigenstates of $\mathcal{H}_2$ with the eigenvalue $J_2$,
\begin{equation}
 \mathcal{H}_2 \lvert 2, S_z \rangle^{\alpha} = J_2 N_0 \lvert 2, S_z \rangle^{\alpha} ,
\end{equation} 
where $S_z=2, 1, 0, -1$ or $-2$.
One of the  $S=1$ states ($\lvert 1, S_z \rangle_a^{\alpha}$) 
has the eigenvalue $-J_2$, and the other two $S=1$ states 
($\lvert 1, S_z \rangle_b^{\alpha}$ and $\lvert 1, S_z \rangle_c^{\alpha}$)
have eigenvalue $0$,
\begin{align}
  \mathcal{H}_2 \lvert 1, S_z \rangle_a^{\alpha} &= - J_2 N_0 \lvert 1, S_z \rangle_a^{\alpha} , \\
  \mathcal{H}_2 \lvert 1, S_z \rangle_b^{\alpha} &=      0 , \\
  \mathcal{H}_2 \lvert 1, S_z \rangle_c^{\alpha} &=      0  ,
\end{align} 
where $S_z= \pm 1$ or $0$.

Two $S=0$ eigenstates ($\lvert 0,0  \rangle_d^{\alpha}$ and $\lvert 0,0  \rangle_e^{\alpha}$)
have eigenvalues $-2J_2$ and $0$, respectively,
\begin{align}
  \mathcal{H}_2 \lvert 0, 0 \rangle_d^{\alpha} &= -2 J_2 N_0 \lvert 0, 0 \rangle_d^{\alpha} , \\
  \mathcal{H}_2 \lvert 0, 0 \rangle_e^{\alpha} &=      0  .
\end{align} 

In Fig.~\ref{figshurikenev0}
the eigenvalues %of four $\alpha$ spins 
of $\mathcal{H}_2 + \mathcal{H}_{\mathrm{Zeeman}}$ are plotted as a function of the external magnetic field 
$h$.
The ground state of $\mathcal{H}_2 + \mathcal{H}_{\mathrm{Zeeman}}$ 
is $\lvert 0,0 \rangle_d^{\alpha}$, 
$\lvert 1,1 \rangle_a^{\alpha}$,   and $\lvert 2,2 \rangle^{\alpha}$ 
for $0<h<J_2$, $J_2 < h < 2 J_2$, and $ h > 2 J_2 $, respectively. 

%%%%%%%%%%%%%%%%%%%%%%%%%%%%%%%%%%%%%%%%%%%%
\section{$2/3$ plateau and $h_4$}
\label{AppendixB}
In this appendix we calculate the critical magnetic field  $h_4$
at which the state changes from the $2/3$-plateau state to 
the state of all spins up.
 The  $2/3$-plateau state is the state in which  four $\alpha$ spins 
(sites 1 - 4 in Fig.~\ref{figshuriken})
  form the $\lvert 1,1 \rangle_a^{\alpha}$ state 
and all $\beta$ spins are up state ($\lvert2,2 \rangle^{\beta}$. 
We consider sites 1 - 8 in Fig.~\ref{figshuriken}.
Then we can write $\mathcal{H}_1$ as
\begin{equation}
 \mathcal{H}_1 = J_1 \sum_{\langle i,j \rangle i=1 \sim 4, j=5 \sim 8}  \mathbf{S}_i \cdot \mathbf{S}_j 
\label{eqappB}
\end{equation}
For example, we consider the term containing $S_5$ in Eq.~(\ref{eqappB}),
\begin{equation}
 J_1 \left[ \frac{1}{2} \left( S_5^-(S_1^+ + S_2^+) +S_5^+(S_1^-+S_2^-)  \right) + S_5^z(S_1^z+S_2^z) \right].
\end{equation} 
Since 
 \begin{equation}
  (S_1^+ + S_2^+)\lvert 1,1 \rangle_a^{\alpha} =0,
 \end{equation}
 and 
\begin{equation}
S_5^+ \lvert 2,2 \rangle^{\beta} =0,
\end{equation}
we can show that the state of the direct product of $\lvert 1,1 \rangle_a^{\alpha}$ 
and all up states of spins $5$ - $8$ 
($\lvert2,2 \rangle^{\beta}$) are
the eigenstates of $\mathcal{H}_1$ and $\mathcal{H}$.
We obtain the energy of this state as
\begin{equation}
 E_{h}^{(h_3 \leq h \leq h_4)} = N_0 (J_1 -J_2 -2h),
\end{equation}
where $N_0$ is the number of the unit cell.
  
  At $h > h_4$ all spins are aligned to the $z$ direction and the energy is
\begin{equation}
  E_{h}^{(h \geq h_4)} = N_0 (2 J_1 + J_2 -3h)  .
\end{equation}
The magnetization jump
 from $M/M_s=2/3$ to $1$ occurs at
$h=h_4$, at which
\begin{equation}
 E_{h}^{(h_3 \leq h \leq h_4)} =E_{h}^{(h \geq h_4)} 
\end{equation} 
We obtain 
\begin{equation}
 h_4 = J_1+2 J_2.
\end{equation}

%%%%%%%%%%%%%%%%%%%%%%%%%%%%%%%%%%%%%%%%%%%%%%%%%%%%%
\section{$0 \leq h <h_1$ and $h_1 < h < h_2$}
\label{AppendixC}
When $h=0$, the true ground state should be the total singlet state of all spins, and the ground state 
at small $h$ might be a complicated state. We do not address the ground state at small $h$ in detail
in this paper.
However, as shown by numerical study\cite{Nakano2014, Nakano2015}, 
in the region of magnetic field $0 < h <h_1$,
$\langle S_i^z \rangle$ is nearly proportional to $h$ for $i \in \beta$, while it is almost zero
for $i \in \alpha$.
We may take a simplified picture that the state for the $\alpha$ spins is approximated as 
the singlet $\vert 00 \rangle_d^{\alpha}$. This approximation is  justified if  $J_1 \ll J_2$, since 
$\vert 00 \rangle_d^{\alpha}$ is the ground state 
for $\mathcal{H}_2 + \mathcal{H}_{\mathrm{Zeeman}}$ for $h<J_2$, as shown in
Fig.~\ref{figshurikenev0}.
 Although  the condition $J_1 \ll J_2$
is not fulfilled in the present case, we treat $\mathcal{H}_1$  as a perturbation.
In the region  $0 < h < h_1$ the system is considered in the state that
the $\alpha$ spins make $\vert 00 \rangle_d^{\alpha}$ and the 
locally excited $\beta$ spin from the singlet state extend over the system  
forming a spin-wave-like state with the repulsive interaction between excitations. 
If there were no interactions between the excitations as in the case at $h=h_4$
discussed in Appendix~\ref{AppendixB}, or if there were attractive interaction between the excitations,
the magnetization jump 
 would occur. 

In the $1/3$-plateau region ($h_1 < h < h_2$), the ground state is approximated by the direct product of  
$\vert 00 \rangle_d^{\alpha}$ for four $\alpha$ spins and 
all $\beta$ spins are aligned up, i.e., $\lvert2,2 \rangle^{\beta}$.
The energy of this state is approximated as
\begin{equation}
 E_{h}^{(h_1 \leq h \leq h_2)} \approx N_0 (-2J_2 -h ).
\end{equation}

%%%%%%%%%%%%%%%%%%%%%%%%%%%%%%%%%%%%%%%%%%%%%%%%%%%%%
\section{$h \gtrsim h_2$}
\label{AppendixD}
In this appendix we show the matrix elements of $\mathcal{H}_1$ between the eigenstates
of $\mathcal{H}_2$ at the magnetic field just above $h_2$.

Using the definition of $\lvert 1 S_z \rangle_a^{\alpha}$, $\lvert 1 S_z \rangle_b^{\alpha}$, 
and $\lvert 1 S_z \rangle_c^{\alpha}$, 
we obtain
\begin{align}
 \mathcal{H}_1 \lvert11\rangle^{\alpha}_a &= J_1 \Bigl[
                           \frac{1}{2\sqrt{2}} \lvert10\rangle^{\alpha}_a  S_a^{+ \beta} \nonumber \\
 &\hspace{5mm}+  \frac{1}{4} \lvert10\rangle^{\alpha}_b             S_b^{+ \beta} \nonumber \\
 &\hspace{5mm}+  \frac{1}{4} \lvert10\rangle^{\alpha}_c            S_c^{+ \beta} \nonumber \\
  &\hspace{5mm}+  \frac{1}{2} \lvert11\rangle^{\alpha}_a           S_a^{z \beta} \nonumber \\
  &\hspace{5mm}+  \frac{1}{2\sqrt{2}} \lvert11\rangle^{\alpha}_b S_b^{z \beta}\nonumber \\
  &\hspace{5mm}+  \frac{1}{2\sqrt{2}} \lvert11\rangle^{\alpha}_c S_c^{z \beta} \Bigr],
\end{align}
\begin{align}
 \mathcal{H}_1 \lvert10\rangle^{\alpha}_a &= J_1 \Bigl[
                           \frac{1}{2\sqrt{2}} \lvert11\rangle^{\alpha}_a  S_a^{- \beta} \nonumber \\
 &\hspace{5mm}+  \frac{1}{4} \lvert11\rangle^{\alpha}_b            S_b^{- \beta}  \nonumber \\
 &\hspace{5mm}+  \frac{1}{4} \lvert11\rangle^{\alpha}_c           S_c^{- \beta} \nonumber \\
  &\hspace{5mm}+  \frac{1}{2} \lvert1-1\rangle^{\alpha}_a         S_a^{+ \beta} \nonumber \\
  &\hspace{5mm}+  \frac{1}{2\sqrt{2}} \lvert1-1\rangle^{\alpha}_b S_b^{+ \beta} \nonumber \\
  &\hspace{5mm}+  \frac{1}{2\sqrt{2}} \lvert1-1\rangle^{\alpha}_c S_c^{+ \beta} \Bigr]
\end{align}
\begin{align}
 \mathcal{H}_1 \lvert1-1\rangle^{\alpha}_a &= J_1 \Bigl[
                           \frac{1}{2\sqrt{2}} \lvert10\rangle^{\alpha}_a  S_a^{- \beta} \nonumber \\
 &\hspace{5mm}+  \frac{1}{4} \lvert10\rangle^{\alpha}_b             S_b^{- \beta} \nonumber \\
 &\hspace{5mm}+  \frac{1}{4} \lvert10\rangle^{\alpha}_c            S_c^{- \beta} \nonumber \\
  &\hspace{5mm}+  \frac{1}{2} \lvert1-1\rangle^{\alpha}_a            S_a^{z\beta} \nonumber \\
  &\hspace{5mm}+  \frac{1}{2\sqrt{2}} \lvert1-1\rangle^{\alpha}_b S_b^{z\beta} \nonumber \\
  &\hspace{5mm}+  \frac{1}{2\sqrt{2}} \lvert1-1\rangle^{\alpha}_c S_c^{z\beta} \Bigr],
\end{align}
where $\mathbf{S}_a^{\beta}$,  $\mathbf{S}_b^{\beta}$,  and $\mathbf{S}_c^{\beta}$
are the spin operators for the $\beta$ spins defined by 
\begin{equation}
 \mathbf{S}_a^{\beta} = \mathbf{S}^{\beta} = \mathbf{S}_5 + \mathbf{S}_6 + \mathbf{S}_7 + \mathbf{S}_8 ,
\end{equation}
\begin{equation}
 \mathbf{S}_b^{\beta} = -\mathbf{S}_5 + \mathbf{S}_6 + \mathbf{S}_7 - \mathbf{S}_8 ,
\end{equation}
and
\begin{equation}
 \mathbf{S}_c^{\beta} =  \mathbf{S}_5 + \mathbf{S}_6 - \mathbf{S}_7 - \mathbf{S}_8 .
\end{equation}
The $z$ component, 
%are defined by 
%\begin{equation}
% S_a^{z\beta} = S_5^z + S_6^z + S_7^z + S_8^z,
%\end{equation}
%\begin{equation}
% S_b^{z\beta} = -S_5^z + S_6^z + S_7^z - S_8^z,
%\end{equation}
%and
%\begin{equation}
% S_c^{z\beta} = S_5^z + S_6^z - S_7^z - S_8^z,
%\end{equation}
%respectively, 
the raising operator, and the lowering operator of 
 $\mathbf{S}_a^{\beta}$,  $\mathbf{S}_b^{\beta}$,  and $\mathbf{S}_c^{\beta}$ are defined as usual,
for example,
\begin{equation}
 S_a^{\pm\beta} = (S_5^x + S_6^x + S_7^x + S_8^x )\pm i (S_5^y + S_6^y + S_7^y + S_8^y ).
\end{equation}

Since $\lvert1 S_z \rangle_b^{\alpha}$, $\lvert1 S_z\rangle_c^{\alpha}$, and 
$\lvert1 -1 \rangle_a^{\alpha}$ have higher energy than
$\lvert1 1 \rangle_a^{\alpha}$ and $\lvert1 0 \rangle_a^{\alpha}$, 
we restrict ourselves in the subspace in $\lvert1 1 \rangle_a^{\alpha}$ and $\lvert1 0 \rangle_a^{\alpha}$
and neglect other states.
Then  $\mathcal{H}_1$ is approximated in the basis of $\lvert 1 1 \rangle_a^{\alpha} \lvert 21 \rangle^{\beta}$ 
and $\lvert1 0 \rangle_a^{\alpha} \lvert 2 2 \rangle^{\beta}$ as
\begin{equation}
 \mathcal{H}_1 \approx \frac{1}{2}J_1 \left( \begin{array}{cc}
 S_a^{z \beta}                             & \frac{1}{\sqrt{2}} S_{a}^{+ \beta} \\
 \frac{1}{\sqrt{2}} S_{a}^{- \beta}  & 0
 \end{array} \right).
\end{equation} 
%%%%%%%%%%%%%%%%
Since
\begin{equation} 
^{\beta}\langle 21 \rvert S_a^{z\beta} \lvert 21 \rangle^{\beta} = 1,
\end{equation}
%%%%%%%%%%%%%%%%%%%%%%%%%%%%%%%%%%%%%%%%%
and 
\begin{equation} 
 ^{\beta}\langle  22 \rvert S_a^{+\beta} \lvert 21 \rangle^{\beta} = 
 ^{\beta}\langle  21 \rvert S_a^{-\beta} \lvert 22 \rangle^{\beta} = 
2,
\end{equation}
we obtain
\begin{equation}
 \mathcal{H}_1 \approx J_1 \left( \begin{array}{cc}
\frac{1}{2}                         & \frac{1}{\sqrt{2}} \\
 \frac{1}{\sqrt{2}}  & 0
 \end{array} \right).
\end{equation} 
In this subspace ($\lvert10\rangle^{\alpha}_a \lvert22\rangle^{\beta}$ 
and $\lvert11\rangle^{\alpha}_a \lvert21\rangle^{\beta}$) the Hamiltonian is approximated as
\begin{equation}
 \mathcal{H}^{(h=h_2+0)} \approx \left(
 \begin{array}{cc}
 \frac{1}{2} J_1 - J_2 - \frac{3}{2} h & \frac{1}{\sqrt{2}} J_1 \\
 \frac{1}{\sqrt{2}} J_1 & -J_2 -h
 \end{array} \right).
 \label{eqapproxh2}
\end{equation}
The eigenvalues of  Eq.~(\ref{eqapproxh2}) are
\begin{equation}
  E_h^{(h=h_2+0)} =   N_0
\left( \frac{1}{4} J_1 -J_2 - \frac{5}{4} h \pm \frac{1}{4} \sqrt{h^2 - 2 J_1 h + 9 J_1^2} \right) .
\end{equation}
We take the minus sign for the square root, since the state with lower energy is realized. 

%%%%%%%%%%%%%%%%%%%%%%%%%%%%%%%%%%%%%%
%%%%%%%%%%%%%%%%%%%%%
%%%%%%%%%%%
%%%%%%%%%%%%%%%%%%%%%%%%%%%%%%%%%%%%%%%%%%%%%%%%%%%%%%%%%%%%%%%%%%%%%%%%%
%% 
%\begin{figure}[tb]
%%
%%\flushleft{ (a) \hfill (b) \hfill \ } \vspace{-0.0cm}\\
%\begin{center}
%%\vspace{1.8cm}
%%\vspace*{0.4cm} 
%\includegraphics[width=0.38\textwidth]{MG-1-04W.EPS} %\\ \vspace {0.5cm}
% %\hspace{0.3cm}
%%
%\end{center}
%%%
%\caption{(color online).
%(a) Magnetization process on square kagome Lattice with $J_2/J_1=1.02$ obtained by the exact diagonalization.
%Black triangles,  green diamonds, blue squares and red circles are obtained in the systems 
%of $N_s=24$, $30$, $36$, and $42$, respectively. (b) Susceptibility obtained numerically in finite systems. }
%\label{fignumerical}
%\end{figure}
%%%%%%%%%%%%%%%%%%%%%%%%%%%%%%%%%%%%%%%%%%%%%%
%

\bibliography{shuriken}
%%%%%%%%%%%%%%%%%%%%%%%%%%%%%%%%%%%%%%%%%
\end{document}